\documentclass{jfm}
\usepackage{subcaption}
\usepackage{graphicx}
\usepackage{newtxtext}
\usepackage{newtxmath}
\usepackage{natbib}
\usepackage[dvipsnames]{xcolor}
\usepackage{hyperref}
\usepackage{amsmath}
\usepackage{mathtools}
\usepackage{newtxtext}
\usepackage{multirow}
\DeclareMathOperator\erf{erf}

\DeclareMathOperator\erfcx{erfcx}
\DeclareMathOperator{\sgn}{sgn}
\newcommand{\defeq}{\vcentcolon=}
\hypersetup{
    colorlinks = true,
    urlcolor   = blue,
    citecolor  = black,
}

\newcommand{\RomanNumeralCaps}[1]
\linenumbers

\usepackage{silence}

\title{On the applicability of the actuator line method for unsteady aerodynamics}

\author{El\'ias Alva\aff{1}
  \corresp{\email{eliasrean@ita.br}},
  Vitor G. Kleine\aff{1},
 \and Andr\'e V.G. Cavalieri\aff{1}}

\affiliation{\aff{1}Divis\~{a}o de Engenharia Aeron\'autica, Instituto Tecnol\'ogico de Aeron\'autica,  S\~{a}o Jos\'{e} dos Campos, SP, 12228-900, Brazil}

\begin{document}

\maketitle

\begin{abstract}

A linear theory for unsteady aerodynamic effects of the actuator line method (ALM) was developed. This theory is validated using two-dimensional ALM simulations, where we compute the unsteady lift generated by the plunging and pitching motion %harmonic transverse oscillation 
of a thin airfoil in uniform flow, comparing the results with Theodorsen's theory. This comparison elucidates the underlying characteristics and limitations of ALM when applied to unsteady aerodynamics. Numerical simulations were conducted across a range of chord lengths and oscillation frequencies. Comparison of ALM results with theoretical predictions shows consistent accuracy, with all Gaussian parameter choices yielding accurate results at low reduced frequencies. Furthermore, the study indicates that selecting a width parameter ratio of $\varepsilon/c$ (the Gaussian width parameter over the chord length) between 0.33 and 0.4 in ALM yields the closest alignment with analytical results across a broader frequency range. Additionally, a proper definition of angle of attack for a pitching airfoil is shown to be important for accurate computations. These findings offer valuable guidance for the application of ALM in unsteady aerodynamics and aeroelasticity.

\end{abstract}

\section{Introduction}

The actuator line method (ALM) is a computational technique that replaces the details of the mesh airfoil geometry by body forces. Initially developed by \cite{sorensen2002numerical}, ALM numerically solves the Navier-Stokes equations without directly resolving the airfoil surface, thus greatly reducing computational cost. Although near-field fidelity is diminished since the airfoil surface is not explicitly resolved, consistent results are obtained for the far field \citep{sorensen2015simulation}. Consequently, ALM has become one of the most widely used methods for modeling wind turbine blades in aerodynamic simulations and has also been employed in simulating propellers, rotorcraft blades, tidal turbines and fixed-wind aircraft lifting surfaces \citep{sorensen2015simulation,kleine2023non,kleine2023simulating}.

In unsteady aerodynamics, the actuator line method offers a broader range of application compared to Theodorsen's theory. Incorporating actuator lines into numerical simulations of the Navier-Stokes system enables the study of more complex configurations. These include multiple surfaces with arbitrary motion, free-stream turbulence and non-uniform flow conditions such as incident sheared flow characteristic of atmospheric boundary layers \citep{troldborg2007actuator,troldborg2009actuator,troldborg2014simple,sorensen2015simulation}, oscillating turbines \citep{cheng2019numerical,arabgolarcheh2022modeling,kleine2022stability} and for aeroelastic studies \citep{churchfield2012numerical,andersen2016statistics,trigaux2024investigation}, among others, where the body forces vary with time. However, to the best of our knowledge, a systematic study of the unsteady aerodynamic response in frequency domain produced by the actuator line method  %the frequency response of the forces under unsteady  conditions 
has not yet been conducted. An accurate computation of unsteady aerodynamic forces across a range of frequencies is essential for studying fluid-structure interactions in both wind energy and aeronautical applications.

This work aims to study the accuracy of ALM for unsteady aerodynamics. To achieve this, ALM is applied to a simplified setup for which an analytical solution exists, allowing for a detailed comparison with Theodorsen's theory. This theory models small oscillations of an airfoil in a uniform flow, providing a robust reference to assess ALM's capabilities and limitations by obtaining its frequency response and juxtaposing it with analytical results.

In this study, a theoretical model for unsteady effects involved in ALM is developed, validated through simulations of unsteady aerodynamics on a thin airfoil. Both pitching and plunging motions are evaluated, and we expect the results from this unsteady motion study will also be applicable to unsteady inflow conditions. Our results are compared with Theodorsen's theory \citep{theodorsen1935general}, which describes how the amplitude and phase of unsteady aerodynamic forces vary with oscillation frequency. Two-dimensional plunging and pitching airfoils with different ALM body force parameters are simulated to assess how closely the numerical results align with Theodorsen’s theory. As ALM is not a closed-form solution, owing to the inherent presence of a free parameter, $\varepsilon/c$, the results aim to provide guidelines for selecting numerical parameters to achieve accurate unsteady aerodynamic flow simulations.

During the writing of this paper, we became aware of an independent study \citep{taschner2024unsteady} proposing a similar theory for unsteady effects in the context of ALM, specifically focusing on pitching motion. In contrast, the present study extends the analysis to include plunging motion and provides a more direct comparison to Theodorsen's theory, elucidating the impact of pitch rate on lift calculations.

This paper is structured as follows: \S 2 provides the theoretical basis and the numerical setup, while in \S 3 presents the linear theory developed results, ALM simulations of pitching and plunging motion in contrast with Theodorsen's theory. Finally, \S 4 addresses the conclusions and discussions.
\section{Methods}
\label{sec:methods}

\subsection{Oscillation of a thin airfoil in uniform flow}

The aerodynamic forces generated by the harmonic motion of an airfoil in uniform flow were analyzed by \cite{theodorsen1935general}. Theodorsen developed an exact solution using the potential flow and the Kutta condition. The unsteady lift ($L$) can be expressed as (\citealt{bisplinghoff2013aeroelasticity})
\begin{equation}
    L = \pi \rho {b}^2 \left[\ddot{h} + U_{\infty} \dot{\alpha_g} - {b} a \ddot \alpha_g\right] + 2 \pi \rho U_{\infty} b C(k)\left[\dot h + U_{\infty} \alpha_g + b\left(\frac{1}{2} - a\right) \dot \alpha_g\right],
 \label{eq:lift_c}
\end{equation}
where $\rho$ is the fluid density, $b$ is the half-chord length, $h$ is the vertical displacement, $U_{\infty}$ is the free-stream velocity, $\alpha_g$ is the geometric angle of attack, $k$ is the reduced frequency based on semi-chord length ($k = \Omega c / 2 U_{\infty}$), $a$ is the distance between the rotation axis and the mid-chord, and $C(k)$ is the Theodorsen's function related to the Hankel functions
\begin{equation}
    C(k) = \frac{H_1^{(2)}(k)}{H_1^{(2)}(k)+iH_0^{(2)}(k)},
\end{equation}
where $H_n^{(2)}$ is a combination of Bessel function of first and second kind $H_n^{(2)}=J_n - iY_n$. This study focuses on the second term of the equation \ref{eq:lift_c}, which corresponds to the circulatory component. Due to the absence of an explicit surface in ALM simulations, we do not expect significant effects from the the first term (apparent mass term). 

The lift coefficient generated by a quasi-steady effect is related to the unsteady lift in equation \ref{eq:lift_c} (considering only the circulatory term) by
\begin{equation}
    C_l(\dot h, \alpha_g, \dot \alpha_g) = C(k) {C_l}_{QS}(\dot h, \alpha_g, \dot \alpha_g),
    \label{eq:theodorsen_qs}
\end{equation}
where ${C_l}_{QS}$ is the quasi-steady lift simplification, which is calculated as
\begin{equation}
    {C_l}_{QS} = 2 \pi \left[\dot h^* + \alpha_g + \left(\frac{1}{2} - a \right) \dot \alpha_g^* \right],
 \label{eq:lift_qs2}
\end{equation}
where $\dot h^* = \dot h / U_{\infty}$, and $\dot \alpha_g^*$ = $\dot \alpha_g$ $b/U_{\infty}$ are the non-dimensional plunge and pitch rate respectively. The relation in equation \ref{eq:theodorsen_qs} indicates that the ratio of the unsteady lift and the quasi-steady lift is governed by Theodorsen's function $C(k)$, which reflects the wake's influence. However, there is an additional simplification, which neglects the pitch rate in equation \ref{eq:lift_qs2},
\begin{equation}
    {C_l}_{SS} = 2 \pi \left[\dot h^* + \alpha_g^*\right],
 \label{eq:lift_qs1}
\end{equation}
commonly used in ALM simulations considering the steady-state conditions, although the neglect of the pitch rate in the equation \ref{eq:lift_qs1} results in a difference in the lift calculation. Later, we will focus on the effects of applying these two simplifications to unsteady aerodynamics.

\subsection{ALM unsteady effects}
\label{sec:2_ALMunsteady}

\subsubsection{Vorticity created by an unsteady lift force convoluted with a 3-d Gaussian function}

The proposed theoretical model is based on the vorticity generated by actuator lines, extending the work of \cite{forsythe2015coupled} and \cite{martinez2019filtered} to consider unsteady effects. For completeness, this section reproduces the derivation initially presented in \cite{kleine2022onstability}. The inviscid vorticity equation, considering unsteady effects, is
\begin{equation}
  \label{eq:vorticityfullequation}
  \frac{\partial\boldsymbol{\omega}}{\partial t}  + \mathbf{u} \cdot \nabla \boldsymbol{\omega} = \boldsymbol{\omega} \cdot \nabla \mathbf{u} + \nabla \times \mathbf{f} .
\end{equation}
We assume an undisturbed velocity in the $z$-direction, $\mathbf{U}=(0,0,U_{\infty})$. If only the lift force, $\mathbf{f}=(0,f_y,0)$, along a line located at $(y,z)=(0,0)$ is considered, the linearization of the vorticity equation considering small disturbances to $\mathbf{U}$ gives rise to the equation \citep{kleine2022onstability}
\begin{equation}
  \label{eq:vorticityfullequation}
  \frac{\partial\boldsymbol{\omega}}{\partial t}  + U \cdot \nabla \boldsymbol{\omega} = \nabla \times \mathbf{f} ,
\end{equation}
which can be written as
\begin{equation}
  \frac{\partial\omega_x}{\partial t} + U_{\infty} \frac{\partial\omega_x}{\partial z} = -\frac{\partial f_y}{\partial z} ,
\end{equation}
\begin{equation}
  \frac{\partial\omega_z}{\partial t} + U_{\infty} \frac{\partial\omega_z}{\partial z} = \frac{\partial f_y}{\partial x} ,
\end{equation}
with the boundary condition $\boldsymbol{\omega} \rightarrow 0$ as $z \rightarrow -\infty$. Assuming a periodic body force with the format $f_y(x,y,z,t)=\widehat{f_y}(x,y,z) e^{i \Omega t}$, the solutions of the first-order linear differential equations are
\begin{equation}
  \label{eq:vortxunsteady}
  \begin{split}
    \omega_x & = \frac{\exp{\left(i \Omega \left(t - \frac{z}{U_{\infty}} \right)\right)}}{U_{\infty}} \int_{-\infty}^{z} -\frac{\partial \widehat{f_y}}{\partial z} \exp{\left(\frac{i \Omega z}{U_{\infty}}\right)} dz \\ & = - \frac{\widehat{f_y}}{U_{\infty}} \exp{(i \Omega t}) +  \frac{\exp{\left(i \Omega \left(t - \frac{z}{U_{\infty}} \right)\right)}}{U_{\infty}} \int_{-\infty}^{z}  \frac{i \Omega}{U_{\infty}} \widehat{f_y} \exp{\left(\frac{i \Omega z}{U_{\infty}}\right)} dz ,
  \end{split}
\end{equation}
\begin{equation}
  \label{eq:vortzunsteady}
  \omega_z = \frac{\exp{\left(i \Omega \left(t - \frac{z}{U_{\infty}} \right)\right)}}{U_{\infty}} \int_{-\infty}^{z} \frac{\partial \widehat{f_y}}{\partial x} \exp{\left(\frac{i \Omega z}{U_{\infty}}\right)} dz .
\end{equation}
We consider an oscillatory force convoluted with the Gaussian kernel such as equation
\begin{equation}
  \label{eq:convolutedforceunsteady}
  \widehat{f_y}(x,y,z)= -\frac{1}{\rho}(\widehat{F_l}\eta)(x) \,  \eta(y) \, \eta(z) ,
\end{equation}
where
\begin{equation}
  \label{eq:eta_d}
  \eta(x) \defeq \frac{1}{\pi^{1/2}\varepsilon} \exp{\left(-\frac{x^{2}}{\varepsilon^{2}}\right)} = \frac{1}{\varepsilon} \eta \left(\frac{x}{\varepsilon} \right), \quad\mathrm{and}\quad \eta(x^*) \defeq \frac{1}{\pi^{1/2}} \exp{\left(-x^{*2}\right)},
\end{equation}
and $F_l(x,t)=\widehat{F_l}(x) e^{i \Omega t}$ is the two dimensional lift force. $\varepsilon$ denotes the smearing parameter, which in numerical simulations should be properly resolved by the chosen grid, and superscript $*$ indicates non-dimensional quantities. This force is written as $F_l = \rho U_{\infty} \Gamma$, which can be interpreted as a linearized form of the Kutta-Joukowski theorem, in the absence of apparent mass effects.

The solutions of equations~\eqref{eq:vortxunsteady} and~\eqref{eq:vortzunsteady} become
\begin{equation}
  \label{eq:vortxunsteady_conv}
  \begin{split}
    \omega_x = & ( \widehat{\Gamma} * \eta)(x) \, \eta(y) \, \eta(z) \exp{(i \Omega t}) \\ 
    & - \frac{i \Omega}{U_{\infty}} (\widehat{\Gamma} * \eta)(x) \,  \eta(y) H_{\varepsilon}\left(\frac{z}{\varepsilon}-i \frac{\Omega \varepsilon}{2 U_{\infty}}\right) \exp{\left(i \Omega \left(t - \frac{z}{U_{\infty}} \right)\right)} \exp{\left( -\frac{\Omega^{2} \varepsilon^{2}}{4 U_{\infty}^{2}} \right)}
  \end{split}
\end{equation}
\begin{equation}
  \label{eq:vortzunsteady_conv}
  \omega_z = - \left(\frac{\partial\widehat{\Gamma}}{\partial x} * \eta\right)(x) \, \eta(y)  H_{\varepsilon}\left(\frac{z}{\varepsilon}-i \frac{\Omega \varepsilon}{2 U_{\infty}}\right) \exp{\left(i \Omega \left(t - \frac{z}{U_{\infty}} \right)\right)} \exp{\left( -\frac{\Omega^{2} \varepsilon^{2}}{4 U_{\infty}^{2}} \right)} ,
\end{equation}
where
\begin{equation}
  H_{\varepsilon}(z)=\frac{\erf(z)+1}{2} .
\end{equation}

From these equations, it is possible to note that a relevant parameter of the problem is the reduced frequency, $k_{\varepsilon}$, defined by $\varepsilon$:
\begin{equation}
  \label{eq:redfreq}
  k_{\varepsilon} = \frac{\Omega \varepsilon}{2 U_{\infty}} ,
\end{equation}
where factor $2$ is used not only because of its explicit occurrence in the formulas, but also due to the convention in the definition of reduced frequency~\citep{bisplinghoff2013aeroelasticity}.
We can define $\omega_x(x,y,z,t)=\widehat{\omega_x}(x,y,z) e^{i \Omega t}$ and $\omega_z(x,y,z,t)=\widehat{\omega_z}(x,y,z) e^{i \Omega t}$.
Non-dimensionalizing by $\varepsilon$ and $U_{\infty}$, the non-dimensional variables are defined as $x^*=x/\varepsilon$, $t^*=tU_{\infty}/\varepsilon$, $\omega^*=\omega\varepsilon/U_{\infty}$, $\Gamma^*=\Gamma/(U_{\infty}\varepsilon)$

\begin{equation}
  \label{eq:vortxunsteady_convke_nd}
  \begin{split}
    \widehat{\omega_x^*} = & ( \widehat{\Gamma^*} * \eta)(x^*) \, \eta(y^*) \,  \eta(z^*) \\ 
    & - i 2 k_{\varepsilon} (\widehat{\Gamma^*} * \eta)(x^*) \,  \eta(y^*) H_{\varepsilon^*}\left(z^*-i k_{\varepsilon} \right) \exp{\left(- i 2 k_{\varepsilon} z^* \right)} \exp{\left( - k_{\varepsilon}^2 \right)}
  \end{split}
\end{equation}
\begin{equation}
  \label{eq:vortzunsteady_convke_nd}
  \widehat{\omega_z^*} = - \left(\frac{\partial\widehat{\Gamma^*}}{\partial x^*} * \eta\right)(x^*) \, \eta(y^*) H_{\varepsilon}\left(z^*-i k_{\varepsilon} \right) \exp{\left(- i 2 k_{\varepsilon} z^*\right)} \exp{\left( -k_{\varepsilon}^2 \right)} .
\end{equation}

\subsubsection{Induced velocity in two-dimensional flow} \label{sec:inducedvelocity}

Now we restrict ourselves to the case of a 2-d flow. From Helmholtz formula~\citep{helmholtz1858integrale}, the velocity induced by vorticity can be written as
\begin{equation}
  \mathbf{u}(\mathbf{x})= \frac{1}{4 \pi} \int \frac{\boldsymbol{\omega}(\mathbf{x'}) \times (\mathbf{x}-\mathbf{x'})}{|\mathbf{x}-\mathbf{x'}|^3} d\mathbf{x'} = \frac{1}{4 \pi} \int \frac{\widehat{\boldsymbol{\omega}}(\mathbf{x'}) \times (\mathbf{x}-\mathbf{x'})}{|\mathbf{x}-\mathbf{x'}|^3} d\mathbf{x'} e^{i \Omega_f t} ,
\end{equation}
hence,
\begin{equation}
  \widehat{\mathbf{u}}(\mathbf{x}) =\frac{1}{4 \pi} \int \frac{\widehat{\boldsymbol{\omega}}(\mathbf{x'}) \times (\mathbf{x}-\mathbf{x'})}{|\mathbf{x}-\mathbf{x'}|^3} d\mathbf{x'} .
\end{equation}
For a two-dimensional flow, restricted to the $y-z$-plane, the vorticity is
\begin{equation}
  \label{eq:vortxunsteady_2d}
  \begin{split}
    \widehat{\omega_x^*} = & \widehat{\Gamma^*}  \, \eta(y^*) \,  \eta(z^*) \\ 
    & - i 2 k_{\varepsilon} \widehat{\Gamma^*} \,  \eta(y^*) H_{\varepsilon}\left(z^*-i k_{\varepsilon} \right) \exp{\left(- i 2 k_{\varepsilon} z^* \right)} \exp{\left( - k_{\varepsilon}^2 \right)}.
  \end{split}
\end{equation}

\begin{figure}
    \begin{subfigure}{0.49\textwidth}
        \includegraphics[width=1\textwidth]{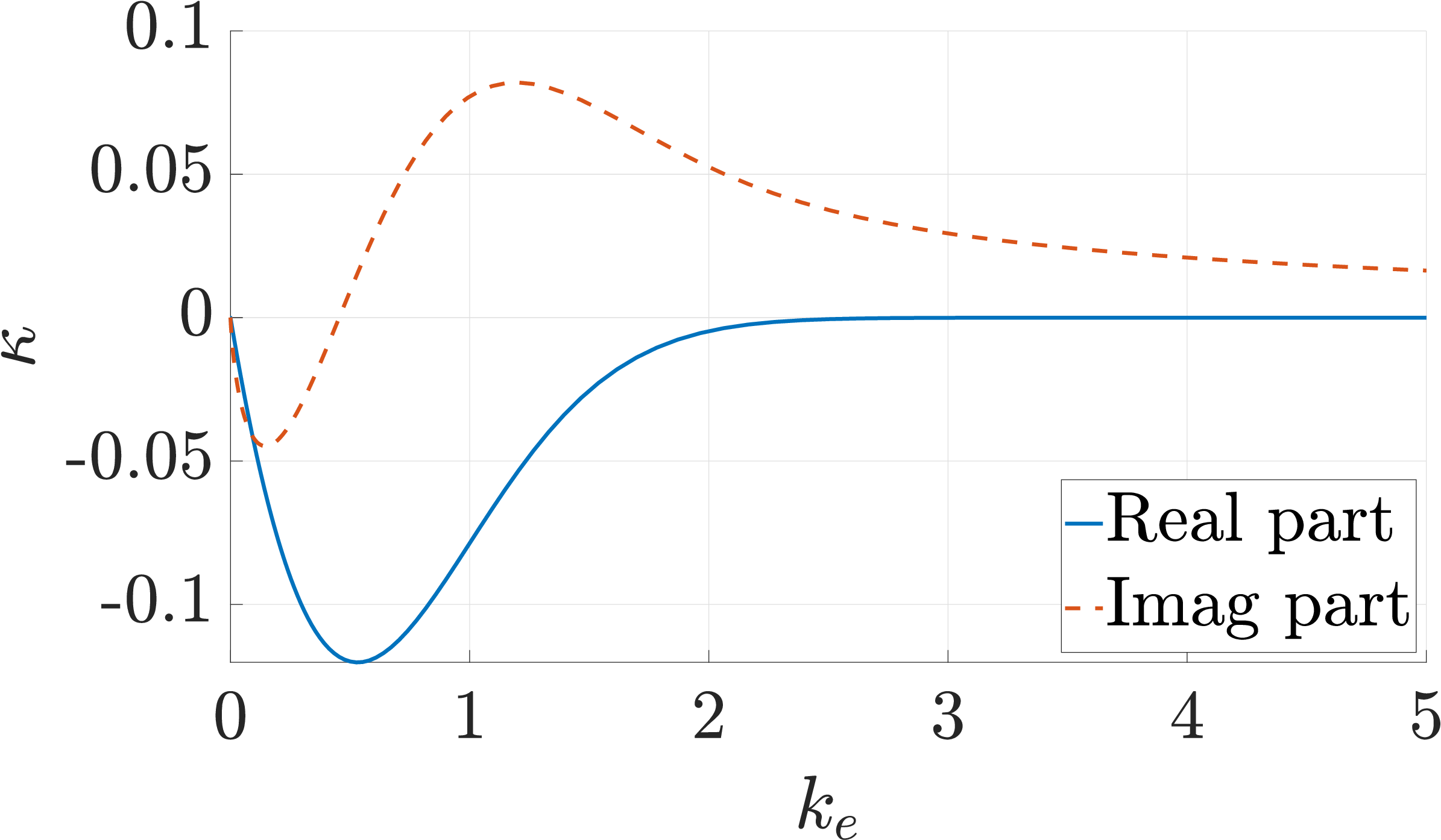}
    \end{subfigure}
    \hfill
    \begin{subfigure}{0.49\textwidth}
        \includegraphics[width=1\textwidth]{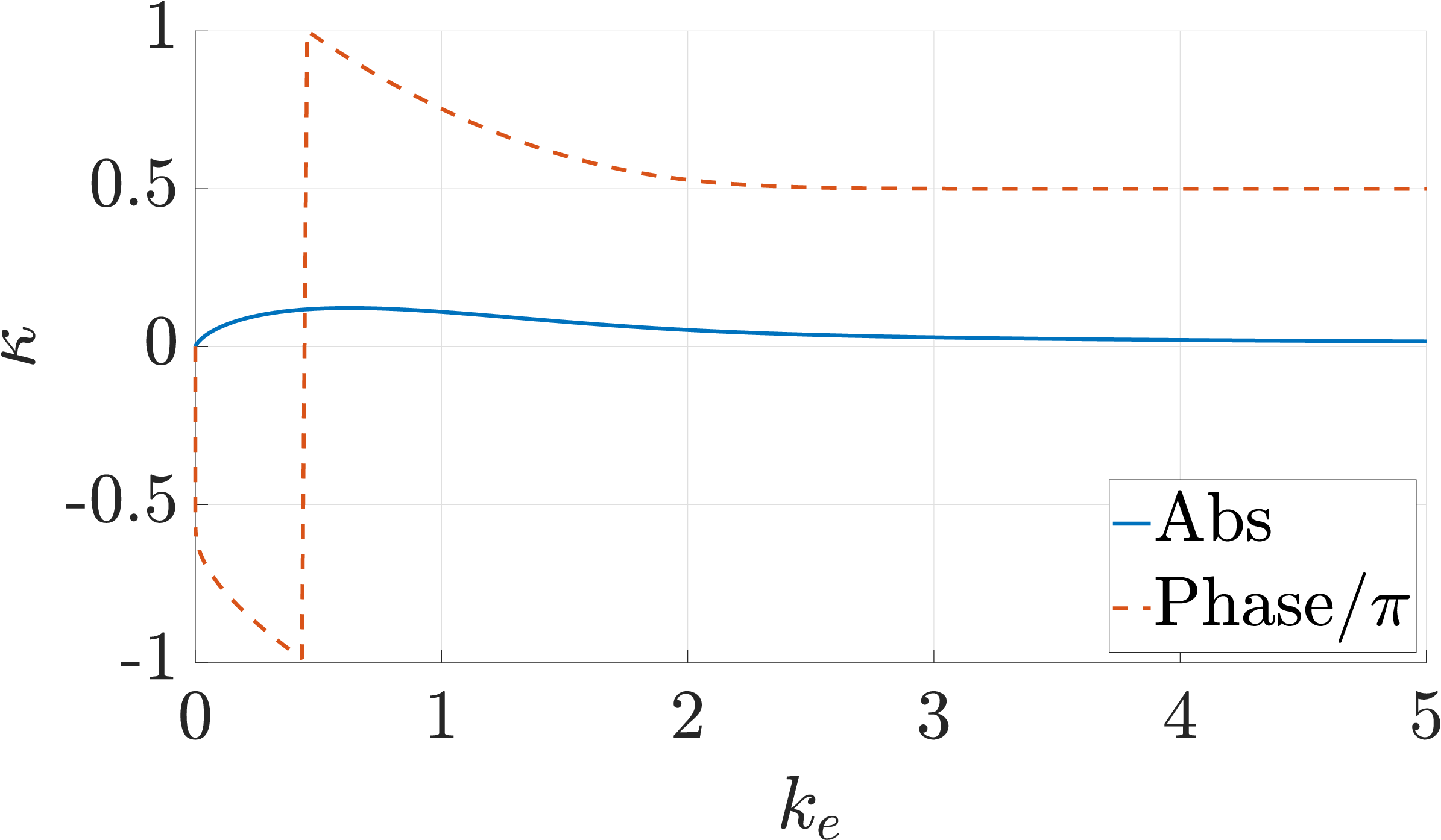}
    \end{subfigure}
    \caption{Function $\kappa$. Ratio between induced velocity and circulation.}
    \label{fig:kappa}
\end{figure}

The first term of equation~\eqref{eq:vortxunsteady_2d} is the bound vortex, which does not induce velocity onto a line located at $(y,z)=(0,0)$, due to symmetry. The induced velocity at the actuator line is then
\begin{equation}
  \begin{split}
      \widehat{u_y}(0,0,0) & = \frac{1}{4 \pi} \int\limits_{-\infty}^{+\infty} \int\limits_{-\infty}^{+\infty} \int\limits_{-\infty}^{+\infty}   \frac{\widehat{\omega_x^*} {z^*}'}{\left({x^*}'^2 + {y^*}'^2 + {z^*}'^2 \right)^{3/2}} d{x^*}' d{y^*}' d{z^*}' \\
      & = - \frac{i 2 k_{\varepsilon} \widehat{{\Gamma^*}} \exp{\left( - k_{\varepsilon}^2 \right)}}{4 \pi} \int\limits_{-\infty}^{+\infty} \int\limits_{-\infty}^{+\infty} \int\limits_{-\infty}^{+\infty} \eta( {y^*}') H_{\varepsilon}\left( {z^*}'-i k_{\varepsilon} \right)\\ & \exp{\left(- i 2 k_{\varepsilon}  {z^*}' \right)} \frac{ {z^*}'}{\left( {x^*}'^2 +  {y^*}'^2 +  {z^*}'^2 \right)^{3/2}} d {x^*}' d {y^*}' d {z^*}' \\
      & = \kappa(k_{\varepsilon}) \widehat{ {\Gamma^*}} ,
  \end{split}
\end{equation}
where
\begin{equation}
  \label{eq:kappa}
  \kappa(k_{\varepsilon}) = - \frac{i 2k_{\varepsilon} \exp(-k_{\varepsilon}^2) }{2 \sqrt{\pi}} \int\limits_{-\infty}^{+\infty} H_{\varepsilon}\left( {z^*}-i k_{\varepsilon} \right) \exp(-i 2 k_{\varepsilon}  {z^*}) \sgn( {z^*}) \erfcx(| {z^*}|) d {z^*} ,
\end{equation}
where $\erfcx(z) = \exp(z^2)\left( 1 - \erf(z) \right)$ is the scaled complementary error function~\citep{johnson2017faddeeva}. The complex function $\kappa$, which is the ratio between the non-dimensional induced velocity and the non-dimensional circulation, is only a function of the reduced frequency $k_{\varepsilon}$. The values of $\kappa(k_{\varepsilon})$, shown in figure~\ref{fig:kappa}, were calculated by integrating equation~\eqref{eq:kappa} in Matlab, using the Faddeeva package~\citep{johnson2017faddeeva} for the error functions.

    \subsubsection{Lift, circulation and angle of attack}

We consider a linear relation between the angle of attack and the airfoil lift coefficient:
\begin{equation}
  C_l = a_0 (\alpha - \alpha_0) ,
\end{equation}
where $\alpha_0$ is the angle for zero lift and $a_0$ is the value of $d C_l/d\alpha$ of the airfoil. In a linear harmonic analysis, $\alpha_0$ only contributes to the steady component of lift. Therefore, the harmonic component can be separated from the steady component:
\begin{equation}
  \widehat{C_l} = a_0 \widehat{\alpha} .
\end{equation}
 {The effective local angle of attack $\alpha$ can be calculated from 
\begin{equation}
    \alpha = \alpha_g + \arctan{\left( \frac{u_{ry}}{u_{rz}}\right)},
\end{equation}
or including the pitch rate term (see section 2.3)
\begin{equation}
    \alpha = \alpha_g + \arctan{\left( \frac{u_{ry}}{u_{rz}}\right)} + b\left(\frac{1}{2}-a \right)\dot\alpha_g,
\end{equation}
where $u_{ry}$ and $u_{rz}$ are the relative velocities. Linearizing the second term
\begin{equation}
    \arctan{\left( \frac{u_{ry}}{u_{rz}} \right)} = \arctan{\left( \frac{u_y + \dot h}{u_{rz}}\right)} \approx \frac{u_y}{U_{\infty}}+\frac{\dot h}{U_{\infty}},
\end{equation}
and linearizing $\alpha$ 
\begin{equation}
    \alpha = \alpha_{QS} + u_y^*.
\end{equation}
% quasi-steady  {linerized} angle of attack $\alpha_{QS}$ 
% \begin{equation}
%    {\alpha_{QS} = \alpha_g + \arctan{\left( \frac{\dot{h}}{U_{\infty}} \right)} + b\left( \frac{1}{2}-a\right)\dot{\alpha_g},}
% \end{equation}
% and the induced velocity of \S~\ref{sec:inducedvelocity}, where $\alpha_{QS}$ does not consider the velocity induced by the shed vorticity {, and $\dot{h}$ is the derivative of the plunging location, when it is applied, more details for the pitch rate term $\dot{\alpha}$ will be exaplining in the \S 2.3}. 
Assuming small induced velocities}
\begin{equation}
  \label{eq:alpha}
  \widehat{\alpha} = \widehat{\alpha}_{QS} + \widehat{ {u_y*}} = \widehat{\alpha}_{QS} + \kappa(k_{\varepsilon}) \widehat{ {\Gamma^*}} .
\end{equation}
Therefore, the lift coefficient can be rewritten as
\begin{equation}
  \label{eq:cl}
  \widehat{C_l} = a_0 \widehat{\alpha}_{QS} + a_0\kappa(k_{\varepsilon}) \widehat{ {\Gamma^*}} = \widehat{C_l}_{QS} + a_0\kappa(k_{\varepsilon}) \widehat{ {\Gamma^*}} ,
\end{equation}
where $\widehat{C_l}_{QS} = a_0 \widehat{\alpha}_{QS}$ is the quasi-steady component of the lift coefficient. The lift force is calculated from:
\begin{equation}
   {F_l} = \frac{1}{2}  {\rho}  {u_{r}}^{2}  {c} C_l ,
\end{equation}
where the relative velocity $ {u_{r}}=\sqrt{ {u_y}^{2}+ {U_{\infty}}^{2}}$ can be linearized to $ {U_{\infty}}$. After linearization
\begin{equation}
   {F_l} =  {\rho}  {U_{\infty} \Gamma} \approx \frac{1}{2}  {\rho}  {U_{\infty}}^{2}  {c} C_l,
\end{equation}
the relation between circulation and lift coefficient in non-dimensional form  becomes
\begin{equation}
  \label{eq:KJ_nd}
   {\Gamma^*} = \frac{ {c^*} C_l}{2}  ,
\end{equation}
where $ {c^*}= {c}/ {\varepsilon}$. Combining equations~\eqref{eq:cl} and~\eqref{eq:KJ_nd}:
\begin{equation}
   \widehat{C_l} = \frac{1}{1-\frac{a_0  {c^*}}{2}\kappa}\widehat{C_l}_{QS} ,
  \label{eq:ratioCl}
\end{equation}
which implies analogous relations for the lift and circulation. We observe that the unsteady lift in the present formulation is a function of the reduced frequency $k_\varepsilon$, and is thus a function of the smearing parameter $\varepsilon$ of the Gaussian kernel; different choices of kernel thus modify the unsteady lift. The absolute value of $\widehat{C_l}/\widehat{C_l}_{QS}$ and its phase difference are shown in figure~\ref{fig:plunge}, which employs the reduced frequency based on the semi-chord, $ {k}$, for direct comparison with Theodorsen's function.

\subsection{Numerical methods}
\label{sec:numerical_methods}

To implement the actuator line method we employ the Dedalus code~\citep{burns2020dedalus}, an open-source framework that enables solving differential equations using spectral methods. We use ALM to compute the lift generated by the harmonic transverse oscillation of a thin airfoil, which may be compared to the present theoretical results, and also to the reference theory by Theodorsen.  {The ALM approach is solved using direct numerical simulations (DNS) of the governing incompressible Navier-Stokes equations, resulting in the following continuity equation}
    \begin{equation}
        \label{eq:continuity}
        \nabla \cdot \mathbf{u} = 0,
    \end{equation}
and the momentum equation in primitive variables (velocity-pressure)
    \begin{equation}
        \label{eq:NS}
        \frac{\partial \mathbf{u}}{\partial t} + \mathbf{u} \cdot \nabla \mathbf{u} = - \frac{1}{\rho} \nabla \mathbf{p} + \nu \nabla ^2 \mathbf{u} + \mathbf{f},
    \end{equation}
where $\rho$ is the fluid density, and $\nu$ is the kinematic viscosity.

In this study, we calculate the body forces $\mathbf{f}$ using the thin airfoil theory, which employs the lift coefficient ($c_l = 2 \pi \alpha$), the free-stream velocity, the fluid density, and the chord length. In the ALM, the angle of attack ($\alpha$) is usually calculated from the local velocity and the geometric angle of attack, $\alpha_g$, according to the steady-state approximation
\begin{equation}
    \alpha = \alpha_g + \arctan \left(\frac{u_y}{ {u_z}}\right),
    \label{eq:alpha_alm_ss}
\end{equation}
where $u_y$ and  {$u_z$} represents the vertical and horizontal components of the relative velocity. In our simulations we sample the velocities at the center of the Gaussian distribution. However, to maintain consistency with Theodorsen's theory, a modified angle of attack
\begin{equation}
    \alpha_m = \alpha_g + \arctan \left(\frac{u_y}{ {u_z}}\right) + b\left(\frac{1}{2} - a\right) \dot \alpha_g,
    \label{eq:alpha_alm_qs}
\end{equation}
with $a = -1/2$, is employed. The differences between cases with and without the pitch rate term ($\dot \alpha $) will be discussed in \S \ref{sec:results}, where it is demonstrated that this term is necessary to achieve results consistent to equation~\ref{eq:theodorsen_qs}. It should be noted that the theory developed in \S \ref{sec:2_ALMunsteady} produces the same ratio of $\widehat{C_l}$ (equation \ref{eq:ratioCl}) for either equation \ref{eq:alpha_alm_ss} or \ref{eq:alpha_alm_qs}, as long as $\widehat{\alpha}_{QS}$ is defined consistently.
The body forces are obtained by convolving the two-dimensional force vector ($\textbf{F}_{2D}$) term \citep{kleine2023non}
    \begin{equation}
       \mathbf{f}^i = \frac{1}{\rho} \mathbf{F}_{2D} \delta (y) \delta ( {z}) ,
       \label{eq:convolution}
    \end{equation}
with a two-dimensional Gaussian kernel
    \begin{equation}
       \mathbf{f} = \mathbf{f}^i * \eta_2 ,
       \label{eq:convolution}
    \end{equation}
    \begin{equation}
       \eta_{2} ( {y,z}) := \frac{1}{\varepsilon^2 \pi} \exp \left(-\frac{ {y^2+z^2}}{\varepsilon^2} \right),
       \label{eq:Gaussian_kernel}
    \end{equation}
where $y$ and  {$z$} denote position, $\eta_{2}$ is the Gaussian kernel, and $\varepsilon$ is the smearing parameter, defining the width parameter of Gaussian function. The value of mesh discretization  {$\Delta y = \Delta z= 4 $} is selected according to \cite{shives2013mesh} (see also \cite{forsting2020generalised} and \cite{kleine2023non}), who concluded that an  {$\varepsilon/\Delta y = \varepsilon/\Delta z$} ratio of approximately 4 minimizes significant errors.

With a fixed value of $\varepsilon = 1$, the chord length of the thin airfoil varies based on $\varepsilon/c$ for each case. The Reynolds number, based on the chord for all simulations, is $Re = c \rho_{\infty} U_{\infty}/{\mu} = 10^5$, where $\rho_{\infty} =1$ is the fluid density, $U_{\infty} = 1$ is the free-stream velocity, and $\mu$ is the dynamic viscosity. An vertical displacement and pitching around a the quarter of chord ($a = -1/2$) are imposed on the thin airfoil, which oscillates according to the following equations
    \begin{equation}
       h = h_0 \sin{\Omega t}, \quad\mathrm{and}\quad \alpha_g = \alpha_0 \sin{\Omega t},
       \label{eq:h}
    \end{equation}
where $h_0 = 0.01 \varepsilon$ and $\alpha_0 = 1^{\circ}$ represent a small amplitude to enable comparison with the linear theory of \S \ref{sec:2_ALMunsteady}. Simulations were conducted over a range of frequencies, and the results are presented in terms of reduced frequency based on the semi-chord ($ {k}$).

    \begin{figure}
        \centering
       %\setkeys{Gin}{width=\linewidth}
       \begin{subfigure}[t]{\textwidth}
            \centering
            \includegraphics[width=0.8\textwidth]{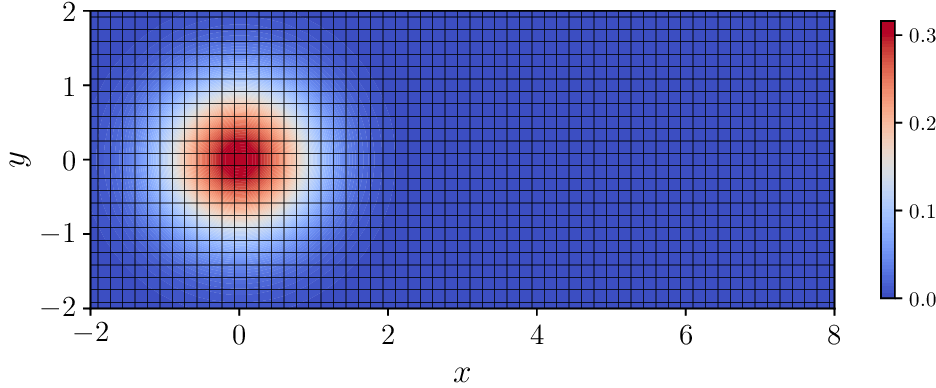}
            \caption{}
            \label{fig:mesh_1}
       \end{subfigure}
       \newline
       \begin{subfigure}[t]{\textwidth}
            \centering
            \includegraphics[width=0.8\textwidth]{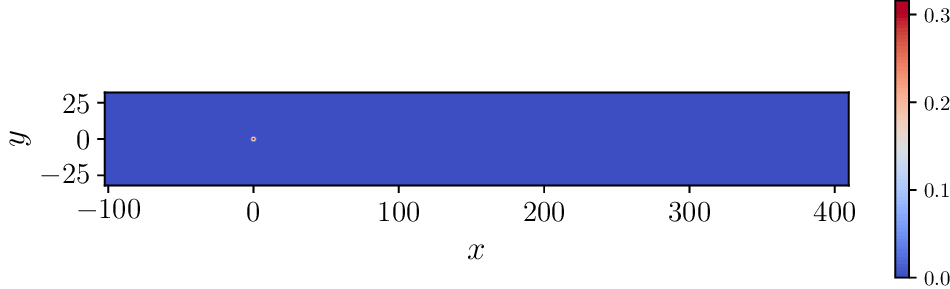}
            \caption{}
            \label{fig:mesh_2}
       \end{subfigure}
       \caption{Illustration of the computational mesh resolution for specific parameters: $\varepsilon$=1.0 (smearing length) and $c=2.5$ (chord length), leading to a ratio of $\varepsilon/c=0.4$. (a) Localized
       view of the mesh, emphasizing the constant discretization and the spatial influence of the Gaussian kernel (represented by the color scale) in the vicinity of the force application point. (b) Overview of the full computational domain, demonstrating the global mesh setup.}
       \label{Fig:mesh_resolution}
    \end{figure}

All parameters are non-dimensionalized by the width parameter of the Gaussian function $\varepsilon$ and the free-stream velocity $U_{\infty}$. The domain dimensions, $L_y$ and $L_{ {z}}$, are chosen to facilitate the calculation of induced velocity for distant wakes at low frequencies. Given the low-frequency values used for thin airfoil oscillation, sufficient space is required for detached vortices to develop. The airfoil is located at the origin of the planar domain ({$h,0$}) for plunging motion, and for pitching motion is located at ($0,0$). For the periodic dimensions, a Fourier basis is used on the  {z-axis} and a sine/cosine basis in the y-axis, which implicitly impose free-slip boundary conditions at the top and bottom domain boundaries. The distance between the inlet ($ {z}_{in}=-102.4$) and the outlet ($y_{out}=409.6$) is $L_{ {z}} = 512$, and the vertical boundary distance is $L_y = 64$, equidistant from the  {z}-axis.  {The mesh resolution has a constant discretization $\Delta z = \Delta y = \varepsilon/4 = 0.25$. Figure \ref{fig:mesh_1} illustrates this resolution within a limited domain, facilitating visualization of the Gaussian function positioned at (0,0) for pitching simulations. Figure \ref{fig:mesh_2} displays the mesh resolution across the entire domain.}

A fringe zone, designed to absorb the wake, is added around the $3/4 $ of $L_{ {z}}$ with a width of $L_{ {z}}/16$. At the outlet, the sponge geometry is carefully defined to prevent wake reflections, enforcing the $y$-component of the velocity to zero and setting the $z$-component to the initial free-stream velocity. We add the fringe as a forcing term $\sigma (\mathbf{u -u_0})$ to the Navier-Stoke equations solver, where $\mathbf{u_0}$ denotes the free-stream velocity and $\sigma$ represents the damping coefficient, which is smoothly varied along x from zero to its maximal value with a tanh function. The accuracy of the fringe is verified by observing that negligible vorticity appears in the left boundary of the domain.  {For the direct numerical simulation, no subgrid was employed. However, to mitigate numerical instability arising from large wavenumbers produced by the highest frequencies in the simulations, a filter provided by the Dedalus framework was utilized. This tool acts as a spatial filter, setting to zero all Fourier coefficients of the velocity solutions associated with wavenumbers larger than a chosen threshold. This effectively introduces artificial dissipation for the energy generated by these larger wavenumbers.}

\section{Results}
\label{sec:results}

A handful of cases were simulated with different ratios $\varepsilon/c = [1/3, 0.4, 0.5, 1, 2, 4]$ and a reduced frequency $ {k}$ between 0.03 and 1.
    \begin{figure}
       \centering
       %\setkeys{Gin}{width=\linewidth}
       \begin{subfigure}[t]{0.7\textwidth}
            \centering
            \includegraphics[width=1\textwidth]{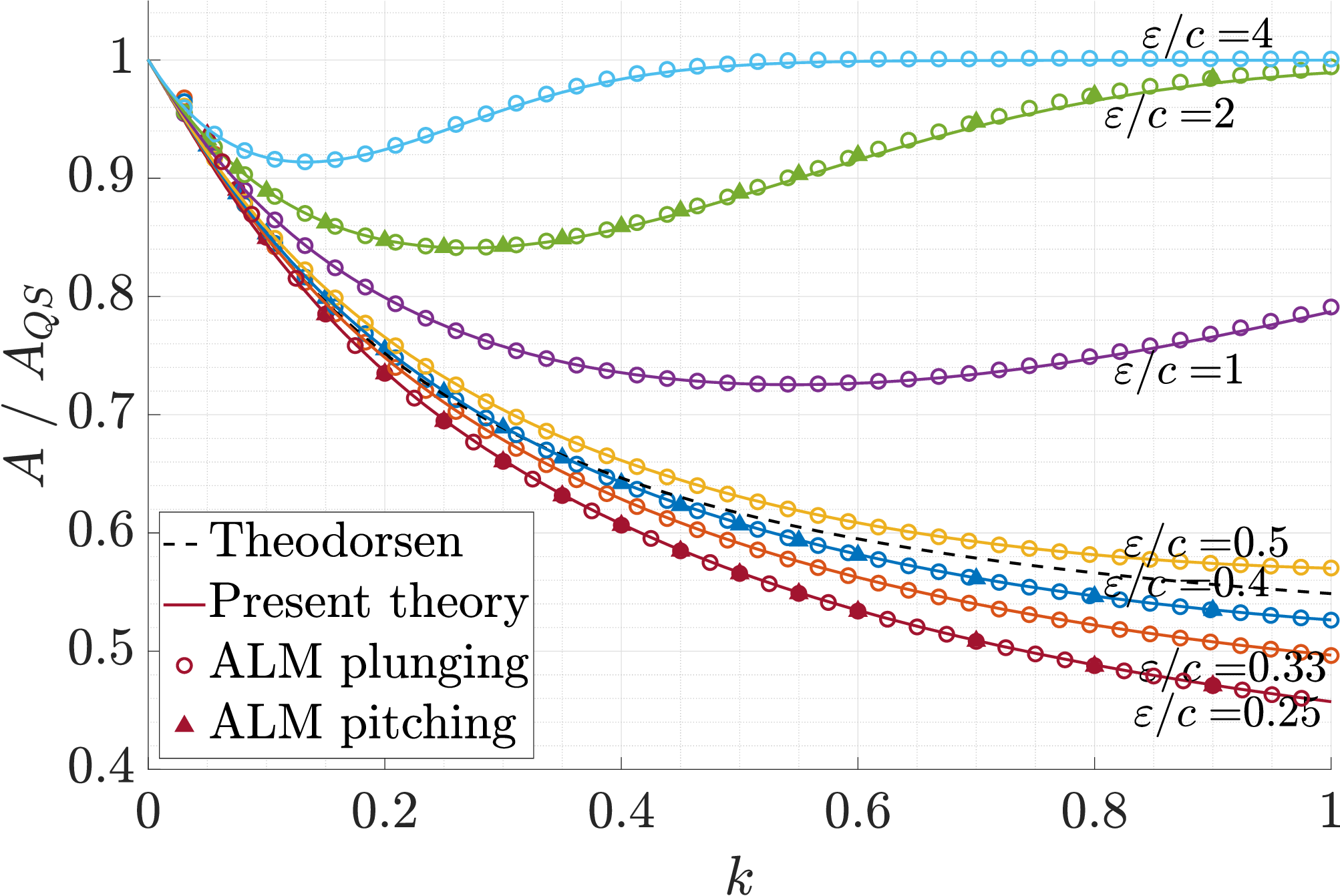}
            \caption{}
       \end{subfigure}
        %\hfill
       \begin{subfigure}[t]{0.7\textwidth}
            \centering
            \includegraphics[width=0.85\textwidth]{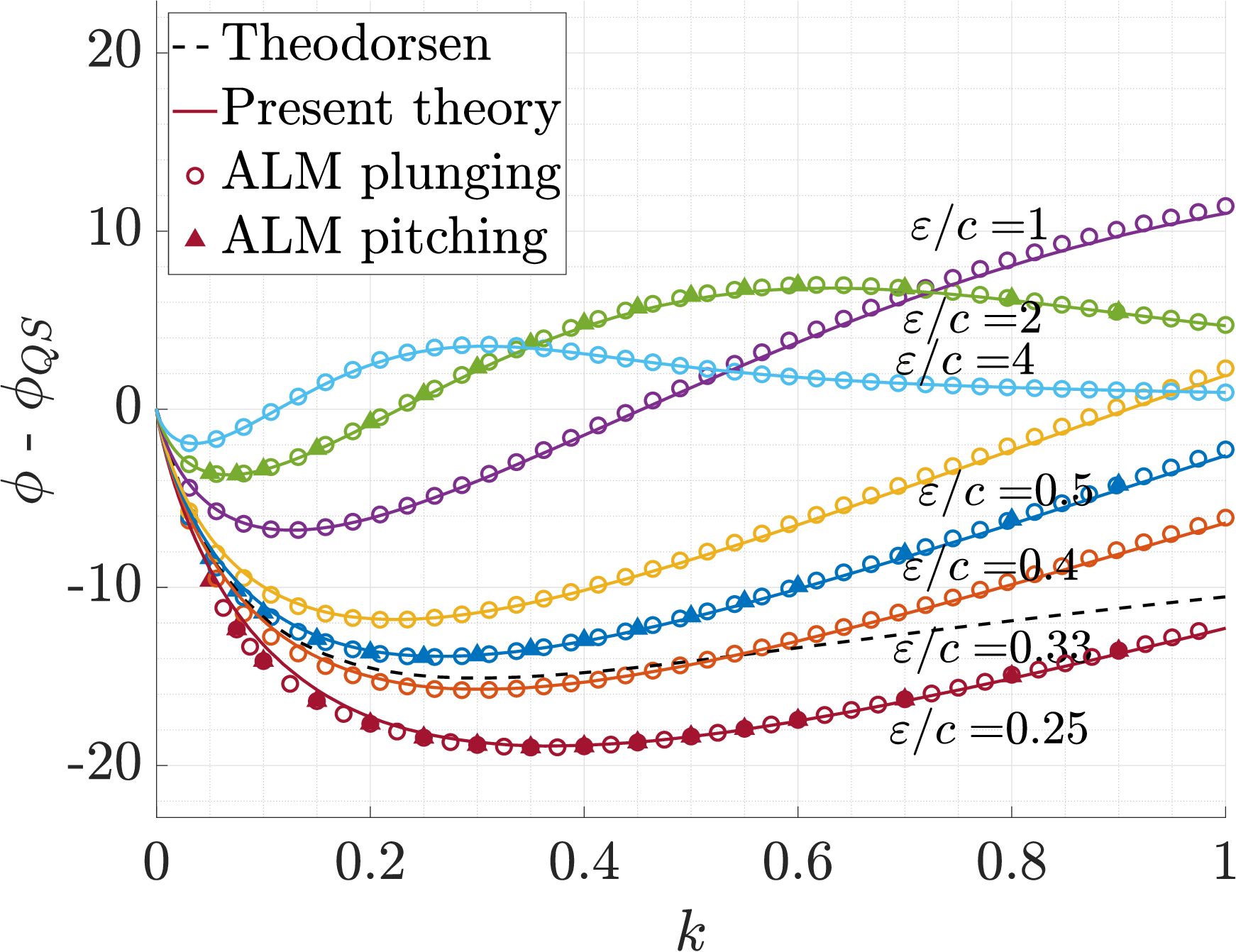}
            \caption{}
       \end{subfigure}
       \caption{Simulation results for plunging motion ($\circ$) and pitching motion ($\blacktriangle$) at different values of the ratio $\varepsilon / c$ and reduced frequency. Solid lines represent the theory of \S \ref{sec:2_ALMunsteady}, while dashed lines show the ratio of the lift coefficient between Theodorsen's theory and the quasi-steady case (considering only the circulatory term). Figures (a) and (b) represent the amplitude and phase difference  {in degrees}, respectively  {. Values of $A_{QS}$ and $\phi_{QS}$ come from equation~\ref{eq:lift_qs2}.}}
       \label{fig:plunge}
    \end{figure} 
Figure \ref{fig:plunge} shows the results for the plunging case. Frame (a) displays the ratio of the amplitude (${A}$) of lift coefficient of the ALM to the quasi-steady case (${A}_{QS}$, from equation~\ref{eq:lift_qs2}), and the phase difference  { ($\phi_{QS}$, also from equation~\ref{eq:lift_qs2})} in frame (b). Solid lines represent the theory of \S \ref{sec:2_ALMunsteady}, while dashed lines show the lift coefficient according to Theodorsen's theory. All simulations yield results consistent with the present theory. As expected, as $ {k}$ approaches zero, all results converge to the quasi-steady model. For higher values of $ {k}$, the results are very dependent on $\varepsilon/c$ and start to diverge from Theodorsen's theory.  { Besides, as figure \ref{fig:plunge} elucidates, the accuracy of the amplitude and phase responses varies, and the choice depends on which type of results are prioritized, i.e. amplitude or phase response.} Until $ {k}=0.5$ the results for the amplitude ratio are close to Theodorsen's theory in the range $\varepsilon/c=0.33$ to $\varepsilon/c=0.5$ and for the phase difference in the range $\varepsilon/c=0.33$ to $\varepsilon/c=0.4$. The agreement between the simulation results and the present theory indicates that the viscous effects are negligible for this Reynolds number.  {Furthermore, an $\varepsilon/c$ value of 0.25, while minimizing near-field airfoil error in steady aerodynamics \citep{martinez2017optimal}, is not the best choice for unsteady cases. %, contradicts the findings of \cite{martinez2017optimal} for steady aerodynamics. 
This indicates that $\varepsilon/c$ values optimized for steady conditions are not suitable for unsteady scenarios, given that $\varepsilon/c=0.25$ produces inferior results compared to the range of 0.33 to 0.5, as illustrated in the figure \ref{fig:plunge}.}

%Additional simulations, focused solely on pitching motion, were performed to validate the linear theory presented in this study. The results of these simulations are consistent with figure \ref{fig:plunge}, if equation \ref{eq:alpha_alm_qs} is used to calculate the angle of attack, with $a = -1/2$.

 {To further validate the presented linear theory, pitching motion was also evaluated under unsteady lift conditions. The primary distinction between pitching and plunging motions lies in the presence of the pitch rate term, which is inherent to pitching but absent in pure plunging motion. This pitch rate generates a linear variation in the normal velocity. Consequently, if this term is neglected, the perturbation velocity cannot be adequately modeled as an arc \citep{leishman2006principles}. Figure \ref{fig:plunge} shows representative results for pitching motion at three values of $\varepsilon/c = 0.25$, 0.4 and 2 in solid triangle markers. These simulations demonstrate consistency with the findings for plunging motion in circle markers, provided that Equation 2.36 is utilized to calculate the angle of attack with $a = -1/2$ to account for the pitch rate.}

\begin{figure}
              \centering
       \setkeys{Gin}{width=\linewidth}
       \begin{subfigure}[t]{0.49\textwidth}
            \includegraphics[width=1\textwidth]{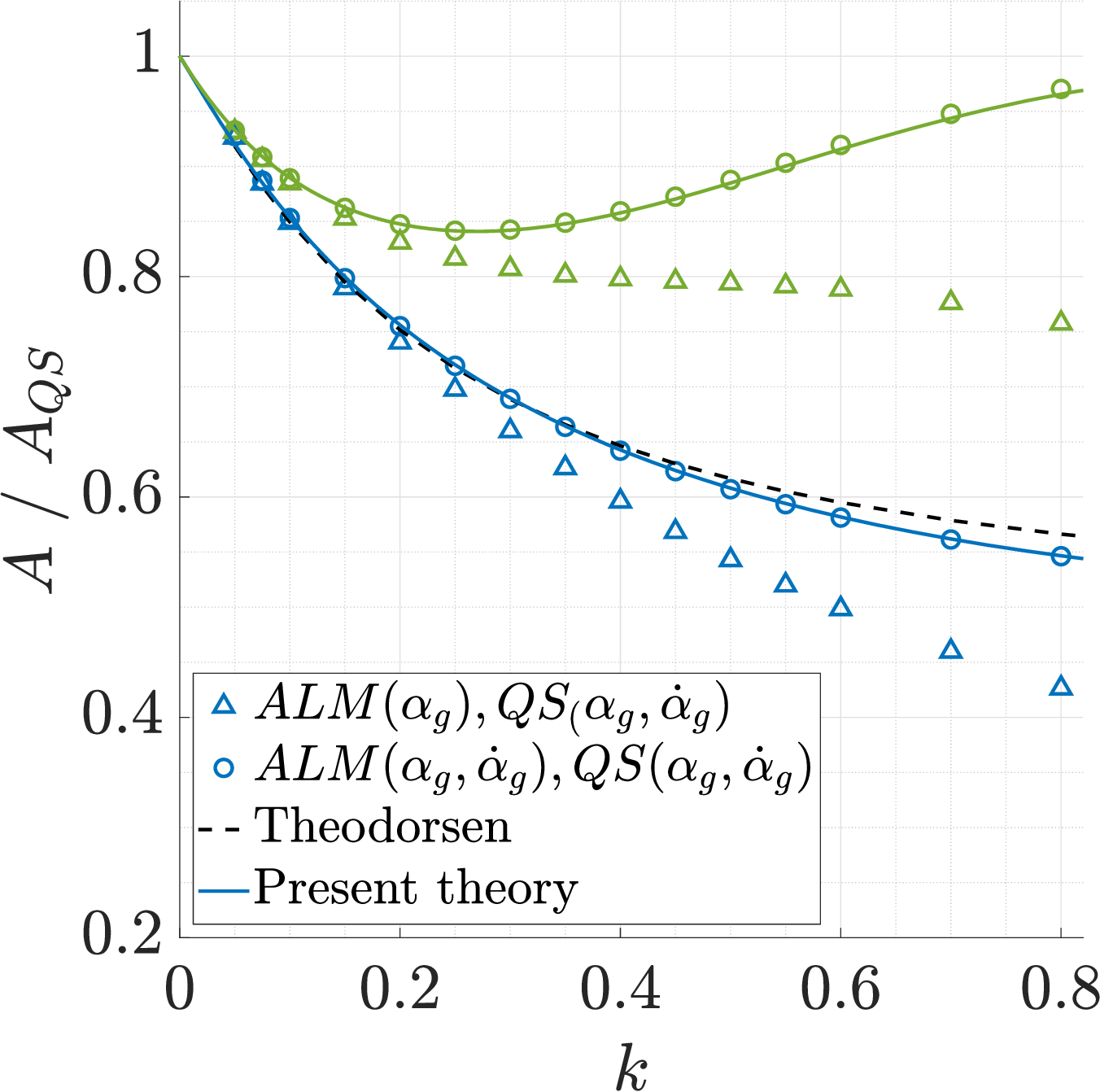}
            \caption{}
       \end{subfigure}
        \hfill
       \begin{subfigure}[t]{0.49\textwidth}
            \includegraphics[width=0.96\textwidth]{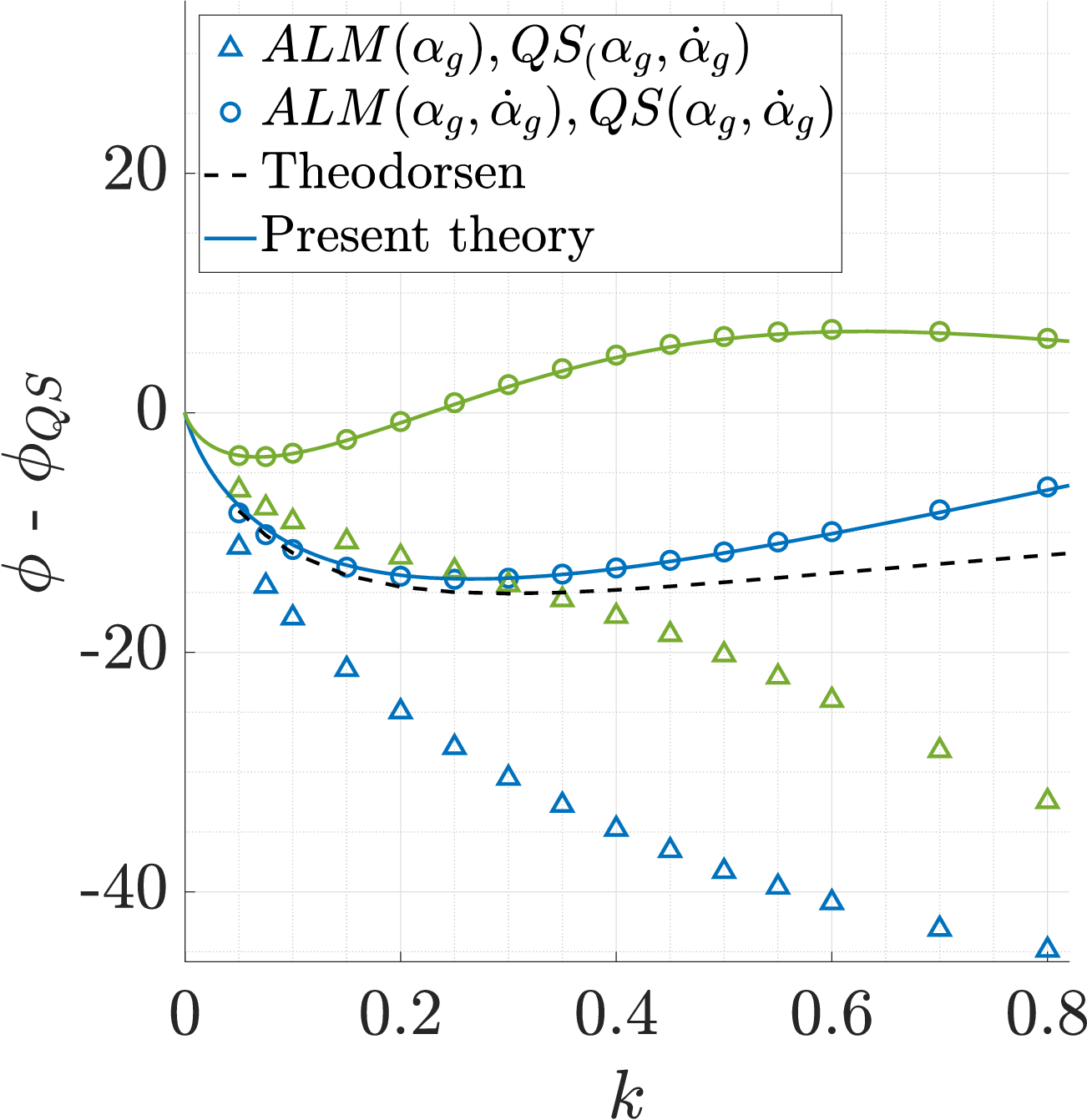}
            \caption{}
       \end{subfigure}
       \caption{Simulation results for pitching motion at a ratio of $\varepsilon / c = 0.4$ in blue markers, and $\varepsilon / c = 2$ in green markers. Circle markers ($\circ$) indicate simulations with damping effects of pitch rate ($\dot \alpha_g$), while triangle markers ($\triangle$) represent simulations without this pitch rate. Solid lines represent the present theory, and dashed lines show the ratio of the lift coefficient between Theodorsen's theory and the quasi-steady case, $|C( {k})|$. Figures (a) and (b) represent the amplitude and phase difference  {in degrees}, respectively.}
       \label{Fig:alpha_alphadot}
    \end{figure}

As we mentioned in \S \ref{sec:numerical_methods} the formulation used to calculate the angle of attack influences the unsteady effects. Figure \ref{Fig:alpha_alphadot} presents simulation results for the case of $\varepsilon/c = 0.4$  {and 2}, comparing the linear theory and ALM simulation results using the quasi-steady formulations, with and without the pitch rate term (equations \ref{eq:alpha_alm_qs} and \ref{eq:alpha_alm_ss}). Both frames show differing results depending on which terms are included in the ALM, if the quasi-steady lift from equation~\ref{eq:lift_qs2} is considered for $A_{QS}$ and $\phi_{QS}$. Consistent results with Theodorsen's theory are obtained when the ALM includes the pitch term. In contrast, if the pitch rate term is not considered, as in equation \ref{eq:alpha_alm_ss}, discrepancies appear in both the amplitude and phase of the results. Therefore, to ensure consistency between Theodorsen's theory and ALM simulations, it is crucial to account for the impact of pitch rate on the lift coefficient in ALM.
\section{  {Conclusions and discussion}}
\label{sec:conclusions}

In the present work, we developed a linear theory for the actuator line method (ALM) applied to unsteady aerodynamics. Numerical  {two-dimensional} simulations were conducted to validate the developed linear theory,  {demonstrating} consistent results across all $\varepsilon/c$ ratio  {employed}. For every $\varepsilon/c$ value, the theory remains consistent within the quasi-steady range, i.e., at low reduced frequencies ($ {k}$). Consistent results between Theodorsen's theory and ALM simulations were found for $\varepsilon/c$ values between  {$\varepsilon/c=0.33$} and  {$\varepsilon/c=0.4$}. For higher reduced frequencies, the linear theory can be used to guide the choice of $\varepsilon/c$ that would lead to more accurate unsteady aerodynamic calculations.

Furthermore, when analysing plunging motion, results proved consistently reliable, as plunging inherently does not involve a pitch rate. Conversely, for pure pitching motion, the inclusion of the pitch rate term in the governing equation is essential. This term enhances the accuracy of the ALM response when compared to Theodorsen's theory, given that the pitch rate leads to a linear variation in the normal velocity and in consequence to the unsteady lift response.

  {It is also pertinent to highlight that employing a different kernel function would undoubtedly yield distinct results. Whether an alternative kernel, such as an anisotropic Gaussian kernel, might provide a more accurate approximation of Theodorsen's theoretical solution compared to the current Gaussian kernel remains to be definitively established. Moreover, although the actuator point position in the present study is fixed at the center of the Gaussian kernel, analysis of differing points could lead to varied results. Nevertheless, the investigation of various kernel functions or alternative actuator point positions represents a valuable direction for future research.}

 {A noteworthy limitation of this study pertains to the assumption of a linear relationship for the airfoil lift coefficient. It is a well-established principle that aerodynamic coefficients, similar to those utilised herein, are typically derived under steady-state conditions. Consequently, the application of these steady aerodynamic coefficients within an unsteady aerodynamic context introduces an inherent limitation, as revealed by the present results. Future studies employing the Actuator Line Method should carry to investigate nonlinear cases or conditions approaching stall.}

%Furthermore, when analyzing the plunging motion, results were consistently reliable, as the plunging does not involve pitch rate due to its nature. Conversely, in the case of pure pitching motion, including the pitch rate term is essential in the equation. %The findings of this work can be valuable for applications in unsteady aerodynamics and aeroelasticity analysis.

%A linear theory for the Actuator Line Method (ALM), as applied to unsteady aerodynamics, was developed in this work. Numerical two-dimensional simulations were conducted to validate the developed linear theory, demonstrating consistent results across all ε/c ratios employed. For every ε/c value, the theory remained consistent within the quasi-steady range, specifically at low reduced frequencies (k). Consistent results between Theodorsen's theory and ALM simulations were observed for ε/c values ranging from ε/c=0.33 to ε/c=0.4. At higher reduced frequencies, the linear theory proves useful in guiding the selection of ε/c to achieve more accurate unsteady aerodynamic calculations.

%Furthermore, when analysing plunging motion, results proved consistently reliable, as plunging inherently does not involve a pitch rate. Conversely, for pure pitching motion, the inclusion of the pitch rate term in the governing equation is essential. This term enhances the accuracy of the ALM response when compared to Theodorsen's theory, given that the pitch rate leads to a linear variation in the normal velocity and in consequence to the unsteady lift response.

\section*{Acknowledgement}
E.A. acknowledges the financial support received from Coordena\c c\~ao de Aperfei\c coamento de Pessoal de N\'ivel Superior (CAPES), under Grants No. 829471/2023-00, and Funda\c c\~ao de Amparo à Pesquisa do Estado de S\~ao Paulo (FAPESP) under Grant No. 2023/12528-1 and grant 2021/11258-5 - Flight and Mobility Innovation Center (FLYMOV).

\bibliographystyle{jfm}
\bibliography{bib}

\end{document}